# Gender and Digital Platform Work During Turbulent Times

Melissa E. Langworthy, Independent Researcher, Calgary, Canada

Yana van der Meulen Rodgers, Rutgers University, Piscataway, NJ, US

**Abstract:** This commentary explores how the platform economy shapes labour market responses during times of crisis, with a focus on gendered experiences. Drawing on cases of economic crisis, natural disasters, and refugee displacement, it examines how digital labour platforms offer flexible work opportunities while also reinforcing existing inequalities. Women face distinct constraints—such as caregiving responsibilities, limited mobility, and economic insecurity—that hinder their employment opportunities and earnings potential. These constraints are more pronounced during crises, when access to stable income and safe working conditions becomes more difficult. While platform work can serve as a lifeline, it is not a guaranteed solution, and its benefits are unevenly distributed. The commentary calls for gender-responsive policies and new research to understand how digital infrastructures mediate labour experiences across different crisis contexts. Such research can inform inclusive strategies that promote resilience and equity in platform-based work, particularly for marginalized and displaced populations.

**Correspondence:** Yana van der Meulen Rodgers, School of Management and Labor Relations, Rutgers University, 94 Rockafeller Road, Piscataway, NJ, 08854, USA. Email yana.rodgers@rutgers.edu

**Keywords:** Gig economy, women, refugees, crisis, precarious work, digital platforms, natural disasters

# 1 | INTRODUCTION

The platform economy operates across global hierarchies, including the global North and South, employers and labourers, and high-skilled and low-skilled workers. It also operates within diverse policy, development, and welfare environments. Platform work is often touted as a potential solution to market challenges like unemployment, as it can provide a critical income stream for individuals who are unemployed or have few options for formal employment. Digital marketplaces also have the potential to provide vulnerable businesses, including those owned by women, with expanded market access and business support services. Despite its apparent ubiquity, the platform economy is not a panacea. Digital labour platforms often expose workers to algorithmic control, income instability, and limited labour protections, contributing to heightened psychological distress and economic precarity (Glavin and Schieman 2022). Gender pay gaps persist regardless of feedback scores, experience, occupational category, working hours, and educational attainment, which suggests that gender inequality is embedded in the operation of platforms (Barzilay and Ben-David, 2016).

The platform economy has a recognized impact on structural arrangements among employers, workers, and consumers, and it is forming new constructions of social life, including organizations of care and social reproduction, precarity, and identity (James, 2024; Dinh & Tienari, 2021; McMillan Cottom, 2020). These new constructions can play a particularly important role during times of crisis and on populations most affected by crisis, especially women. Women often face gendered constraints that shape their participation in the platform economy, including their role as care providers, limited access to financial services, mobility restrictions, and heightened exposure to harassment in certain types of platform work. These intersecting challenges contribute to unequal outcomes and reinforce structural barriers that affect how women engage with digital labour platforms during times of crisis.

In this commentary we reflect on the platform economy and its implications for shifting labour markets, organizations, and employment practices in turbulent times, with special attention to how men and women sometimes experience these shifts differently due to the structure of constraints in which they live and work. This commentary is situated within the context of a growing polycrisis (Wood et al. 2024), where overlapping disruptions such as economic instability, climate-related disasters, and forced displacement converge to reshape labour markets and social life. These intersecting crises intensify precarity and expose structural inequalities, particularly in how men and women engage with and are affected by platform-based work.

## 2 | BACKGROUND

The platform economy refers to a system of economic exchange in which digital platforms serve as intermediaries that facilitate transactions between service providers and consumers, often through algorithmic matching and user ratings. Workers can take on jobs in a piecemeal fashion, usually without a guarantee of additional work. Platform work can be performed in a location (such as driving, delivery, and care), online (such as graphic design, photo editing, coding, writing), or both (such as sex work) (Datta et al., 2023; Rand & Stegeman, 2023). Platform-based work spans a wide spectrum—from highly skilled professionals engaged in complex, long-term projects to lower-skilled workers performing short, repetitive tasks with minimal entry barriers. Platform work is a form of gig work, and the gig economy also includes other types of informal, freelance, and contract-based labour that exist outside of digital platforms (Caza et al. 2022). While digital platforms have reshaped the visibility and scale of gig work, the underlying labour dynamics—particularly for women—are not new. Historically, women have been concentrated in informal, flexible, and low-paid work, from domestic labour to piece-rate manufacturing. The gig economy represents a technological evolution of these longstanding patterns, rather than a wholly novel structure. Recognizing this continuity helps distinguish between the material conditions of gig work and the structural forces, including platform technologies, that shape its current form.

The platform economy is large and growing quickly. The platform economy employs between 154 million and 435 million workers globally, comprising between 4.4% and 12.5% of the global labour force (Datta et al., 2023). Given its growing share of the global workforce, digital gig work is reshaping the labour landscape in profound ways. The rapid growth of platform-based work is closely linked to broader processes of capitalist restructuring, including the rise of flexible, financialized, and fissured labor markets that have intensified precarity and fragmented employment relationships (Peck and Theodore, 2012). While related concepts such as "nonstandard," "contingent," and "precarious" work offer important insights, our focus on platform-based labor highlights how digital infrastructures mediate access to work and shape gendered experiences in new ways.

The global nature of platform-based work can reduce geographic skill mismatches (for example, by offering more opportunities for workers in rural areas or in countries with fewer economic opportunities in a particular field). Individuals with access to platform work are, therefore, less likely to apply for unemployment benefits and better able to pay their bills (Fos et al., 2025). Proponents of platform-based work also cite its potential to bring increased inclusivity to a workforce (for example, by offering work-from-home options for persons with disabilities and care responsibilities) and to provide work opportunities to job seekers outside cities.

Critics of digital labour platforms have objected to the way in which these platforms have exploited their unique position to create significant power imbalances, leveraging their control over information and user interactions to disadvantage workers and consumers (Calo and Rosenblat, 2017; Rosenblat et al., 2017). They argue that the high profits of platform-based companies come primarily from the labour of underpaid workers. Through the narratives of social good and entrepreneurship, workers are being manipulated into taking on a greater share of the liability of learning how to use and implement digital technologies. Workers are further disadvantaged because rating systems are often one-way in that they only give clients the ability to see data about workers and not the opposite (Rosenblat et al., 2017; Ticona and Mateescu, 2018). Thus, workers have little agency in the work they take on and the clients they serve. They may even become dispossessed of their own living space, as in the case of people who rent out their entire home through Airbnb as a means to survive in cities with high costs of living (Maier & Gilchrist, 2022). Even when some aspects of a platform allow workers to set some of their terms with clients, the workers remain subordinated to the authority of platforms (Wood and Lehdonvirta, 2021). The inclusivity of platform-based employment also remains a point of contention. For example, people with disabilities and care responsibilities may not be choosing platform-based work due to the flexibility such jobs provide; rather, they could be being tracked into such work due to labour market discrimination and limited opportunities to obtain standard jobs (Schur and Kruse, 2021).

Globally, women are disproportionately engaged in online gig work compared to their representation in the overall labour market (42% versus 39.7% in 2021) (Datta et al., 2023). Women are particularly well represented in the digital platform economy in the Middle East and North Africa, where 56% of online freelancers are women; in comparison, only 18% of standard jobs in the service sector and 29% of other informal sector jobs are held by women (Datta et al, 2023). Thus, there is evidence to show that the platform economy is providing some opportunities to escape discriminatory workplaces, especially for women. However, digital labour platforms are characterized by substantial gender disparities related to occupational segregation (e.g., women in lower-skilled tasks and men in higher-skilled tasks) and pay. For example, Cook et al. (2021) find that women Uber drivers earn 7% less than men drivers, a difference that is primarily explained by gender differences in learning by doing through using the app, constraints around driving locations related to safety and the need to work close to one's family, and driving speed (with men's faster driving speeds allowing them to complete more rides). In its global reach, the platform economy can exacerbate discrepancies in ICT skills and infrastructure access in ways that relegate women to lower-value 'on-demand' work that replicates their existing skills in caretaking, beauty services, and domestic work, rather than offering opportunities for higher-skill work (Hunt et al. 2017).

The global nature of platform-based work may compromise its capacity to provide fair working conditions. Most online labour demand originates from urban areas in high-income countries, while the supply of platform workers mainly comes from East Europe, South Asia, and the Philippines (Braesemann et al., 2022). This globalized system implies that workers often face competition from workers in other countries with different payment expectations and protections. This fragmentation further introduces elements reminiscent of the offshoring of labour and 'race to the bottom' for labour costs that have defined discussions of globalization. In their home countries, platform-based workers are often categorized as self-employed or independent contractors rather than employees, leading to a lack of formal employment benefits such as minimum wages, overtime payments, health insurance, and paid time off.

## 3 | CRISIS AND DIGITAL LABOUR PLATFORMS

Digital labour platforms have become a major source of employment creation during turbulent times. In this section we focus on three particular crisis situations: economic crises, natural disasters, and refugee crises. Our aim is to draw clear awareness of how the platform economy has come to shape new constructions of work and life in ways that differ between men and women.

During economic crises, when traditional industries reduce hiring or lay off employees, digital platforms allow individuals to find remote and location-based work, providing services and completing projects. The platform economy can thus offer a stopgap solution for displaced workers, enabling them to monetize their skills and assets. A case in point is the COVID-19 pandemic, which was not only a health crisis but also an economic crisis when lockdowns and restrictions led to enormous losses in standard jobs. Many people turned to platform work as a coping mechanism. However, platform work did not necessarily guarantee a steady source of income during the early months when lockdowns wreaked havoc for some platform-based workers, especially drivers, who faced a precipitous decline in demand for their services. In India, 90% of ride-sharing drivers and 75% of delivery workers experienced large earnings losses early in the pandemic (Flourish Ventures, 2020). In many countries, women experienced a disproportionate burden imposed by lockdown policies, especially in terms of relatively greater unpaid care work burdens, which acted as a constraint on women's participation and productivity in the labour market (Bluedorn et al., 2023; Kabeer et al., 2021).

Once lockdowns started to ease, digital platforms provided some relief for unemployed workers, but the growth in platform jobs was gendered: sectors predominantly occupied by women, such as care work and wellness services, saw a decline, whereas sectors typically dominated by men, like driving and food delivery, witnessed a meaningful increase in demand (Blanchard and Hunt, 2022). This uneven demand growth across sectors drew more women into transportation services and contributed to a shift in the composition of platform workers toward more women. In the U.S., women comprised 44% of transportation and delivery platform workers in 2021, up from 36% just two years earlier (Garin et al., 2023). Gender differences among platform workers during economic crises have also been documented in the case of coping mechanisms. For example, in Indonesia, women platform workers were more likely to rely on their personal savings, reduce spending on necessities, and apply for government assistance rather than seek formal loans for which they believed they would not qualify. Men, on the other hand, more often opted for formal loans and the liquidation of business assets (Elhan-Kayalar et al., 2022).

Another case we consider is natural disasters. After natural calamities such as hurricanes, earthquakes, and floods, digital platforms can connect workers with short-term jobs like debris removal, emergency repairs, and logistics. Digital labour platforms can also help to speed recovery efforts while offering immediate income to those affected. However, the extent to which digital platforms are helpful in the aftermath of natural disasters can vary markedly by gender. Men are often more likely to engage in debris removal and emergency repairs, which can provide immediate income but also expose them to higher risks of injury and health hazards, while women may face greater challenges in accessing these opportunities due to safety concerns and caregiving responsibilities (Cortés and Pan, 2018; Lordan and Pischke, 2022). Women often bear the brunt of increased domestic labour and caregiving duties during and after disasters, which can limit their availability for platform-based work (Fatema et al., 2023).

Economic insecurity amplifies the impact of natural disasters on women. Gendered divisions of labour—both at home and in the workforce—leave women with less control over resources, limiting their ability to prepare for and recover from crises. Pre-existing poverty, informal employment, limited land ownership, and heavy domestic responsibilities deepen this vulnerability (Enarson, 2000). These disparities highlight the need for gender-responsive policies and targeted support. Expanding equitable access to platform-based work and tailored recovery programs can help close these gaps and strengthen community resilience.

The final case is refugee crises. When conflict forces people to flee their homes for other countries, the scaffolding of support structures and success outcomes can change radically. Refugees may also face barriers to traditional employment due to legal restrictions or credential recognition. Digital labour platforms have the potential to provide accessible, flexible work opportunities, thus creating income opportunities for refugees while integrating them into local economies. Online platforms can help prevent wage theft and ensure prompt payment, thus providing an income stream to refugees despite their being displaced (Hunt et al., 2018). Platform-based work can also help women refugees who have caregiving responsibilities. For example, Syrian women refugees in Jordan were able to overcome barriers to mobility by engaging in crowd work, completing tasks commissioned online (Hunt et al., 2018).

However, refugees face several challenges in leveraging these opportunities, including limited internet connectivity, digital literacy, lack of access to financial services, and legal uncertainties around work permits (Easton-Calabria and Hackl, 2023). These limitations placed binding constraints on the ability of Syrian women refugees to engage in platform work in Jordan (UN Women, 2020). Concerns about privacy and the lack of social protections can further hinder the participation of refugees in the platform economy. The work is often more precarious for refugees and migrants than it is for a country's citizens. In high-income countries, migrant workers are disproportionately represented in platform work, where their residence status makes them particularly vulnerable to low pay and exclusion from social safety nets (van Doorn et al., 2023). Women refugees and migrants who use digital platforms to find care jobs are particularly vulnerable to sexual harassment and physical abuse given that the work takes place in the private sphere, protective legislation is usually poor or nonexistent, and fears of deportation prevent the women from speaking out. Addressing these barriers through supportive policies and infrastructure improvements is crucial for shifting platform-based work from a 'quick fix' during a humanitarian crises to a longer term opportunity for sustainable development (Easton-Calabria and Hackl, 2023).

## 4 | CONCLUSION AND FUTURE RESEARCH

The platform economy is a site where contradictory forces converge. On one hand, digital platforms have enabled greater labor market participation for women, often supported by progressive policy frameworks and shifting norms around gender equality. On the other hand, these same platforms reproduce and even deepen gendered labor market segmentation. Women are overrepresented in low-paid, flexible, and feminized service roles, while still bearing the brunt of unpaid reproductive labor. This duality reflects a broader reconfiguration of the relationship between the productive and gender orders in contemporary capitalism, where new forms of labor flexibility coexist with persistent inequalities in the sexual division of labor (Mills 2016).

In the context of crisis and upheaval, digital labour platforms can provide a meaningful stopgap measure for vulnerable workers to maintain an income stream. However, governments have a responsibility to providing platform-based workers with access to the resources and support they need to maintain their economic activities and resilience during and after a crisis. With their caregiving responsibilities and pre-existing disparities in the labour market, women are disproportionately affected by crisis and face more hurdles in adapting to the new normal. To help foster self-sufficiency, platform workers – especially women – need support, training, and access to finance. Competition in online markets may, in some cases, require policy interventions to foster a more business-friendly environment (Elhan-Kayalar et al. 2022).

We close with suggestions for future research in five key areas to better understand how gender inequalities are shaped and sustained, particularly in times of crisis. First, scholars should examine how platform work influences gender disparities in access, income, occupational segmentation, and opportunities for skill development, especially in comparison to traditional labor markets. Second, more attention is needed on how the platform economy interacts with the unequal division of reproductive labor, particularly in contexts where women must limit their work to specific times or locations to accommodate caregiving responsibilities—constraints that often intensify during crises.

Third, new research should explore the diversity of platform work itself—ranging from high-skill digital tasks to low-wage, in-person services—and how gender dynamics vary across these segments based on factors like autonomy, regulations, and whether the work is a primary or secondary source of income. Fourth, the role of the platform economy during crises—whether economic, environmental, or humanitarian—deserves closer scrutiny, especially in terms of how it may amplify or mitigate gender inequalities and how state responses shape these outcomes. Fifth, an intersectional approach is essential to understand how gender interacts with other structural inequalities—such as race, migration status, age, and caregiving burdens—particularly in under-researched contexts in the Global South. All told, this new research will help to inform more inclusive and equitable policy responses to the challenges and opportunities of platform-based work.